\documentclass{iaus}
\usepackage{graphicx}

\title[IAU Symposium 262.~~Stellar Populations -- Planning for the Next Decade] 
{Metallicity Mapping with $gri$ Photometry: \\ The Virgo Overdensity
and the \\ Halos of the Galaxy}

\author[Beers et al.]   
{Timothy C. Beers$^{1,2}$, Deokkeun An$^3$, Jennifer A. Johnson$^4$, Marc H. Pinsonneault$^4$,
Donald M. Terndrup$^4$, Franck Delahaye$^5$, Young Sun Lee$^{1,2}$, Thomas Masseron$^4$,
Daniela Carollo$^{6,7}$, and Brian Yanny$^8$}

\affiliation{$^1$Department of Physics \& Astronomy, Michigan State
University, \break  email: beers@pa.msu.edu \\[\affilskip]
$^2$Joint Institute for Nuclear Astrophysics\\[\affilskip]
$^3$IPAC, Caltech\\[\affilskip]
$^4$Department of Astronomy, Ohio State University\\[\affilskip]
$^5$CEA-Saclay, DSM/IRFU/Service d'Astrophysique, France \\[\affilskip]
$^6$Research School of Astronomy \& Astrophysics, ANU, Australia\\[\affilskip]
$^7$INAF, Osservatorio Astronomico di Torino,\\[\affilskip]
$^8$FermiLab}

\pubyear{2009}
\volume{262}  
\jname{IAU Symposium 262.~~Stellar Populations -- Planning for the Next Decade}
\editors{G. Bruzual \& S. Charlot, eds.}
\begin{document}

\maketitle

\begin{abstract}

We describe the methodology required for estimation of photometric
estimates of metallicity based on the SDSS $gri$ passbands, which can be
used to probe the properties of main-sequence stars beyond $\sim$ 10 kpc,
complementing studies of nearby stars from more metallicity-sensitive
color indices that involve the $u$ passband. As a first application of
this approach, we determine photometric metal abundance estimates for
individual main-sequence stars in the Virgo Overdensity, which covers
almost 1000 deg$^2$ on the sky, based on a calibration of the metallicity
sensitivity of stellar isochrones in the $gri$ filter passbands using
field stars with well-determined spectroscopic metal abundances.
Despite the low precision of the method for individual stars, internal
errors of $\sigma_{\rm [Fe/H]} \sim 0.1$ dex can be achieved for bulk
stellar populations. The global metal abundance of the Virgo
Overdensity determined in this way is $\langle$[Fe/H]$\rangle$ $= -2.0
\pm 0.1$ (internal) $\pm 0.5$ (systematic), from photometric
measurements of 0.7 million stars with heliocentric distances from
$\sim 10$ kpc to $\sim 20$ kpc. A preliminary metallicity map, based
on results for 2.9 million stars in the northern SDSS DR-7
footprint, exhibits a shift to lower metallicities as one proceeds
from the inner- to the outer-halo population, consistent with recent
interpretation of the kinematics of local samples of stars with
spectroscopically available metallicity estimates and full space
motions. 


\end{abstract}

\firstsection 

\section{Introduction}

The structure, chemistry, and kinematics of the stellar halo of the
Galaxy, with its predominantly old and metal-poor populations,
collectively preserve a detailed record of our Galaxy's formation in
the early universe. Thanks to large-area surveys such as the Sloan
Digital Sky Survey (SDSS; \cite[York et al. 2000]{York00}), recent studies have
revealed that the halo is marked by numerous stellar substructures.
The presence of these lumpy and complex substructures (both in real
space and in phase space) are in qualitative agreement with models for
the formation of the stellar halo through the hierarchical merging and
accretion of low-mass sub-halos (e.g., \cite[Bullock \& Johnston
2005]{Buljoh05}). Among the various
substructures discovered to date, the Virgo Overdensity (VOD) is one
of the most striking. It was discovered as a stellar overdensity of
main-sequence stars in SDSS; star counts in the region toward Virgo
are enhanced by a factor of two above the background stellar
distribution (\cite[Juri{\'c} et al. 2008]{Jur08}). The overdensity seems to be associated
with clumps of RR Lyrae stars and turn-off stars, but it is less
likely to be connected with the leading tidal tail of the Sagittarius
dwarf galaxy (e.g., \cite[Newberg et al. 2007]{New07}, and references
therein). At present, metallicity estimation from
broadband photometry is the only practical means of obtaining metal
abundances for a large number of faint objects such as those in the
VOD. Such methods are based on the relative sensitivity of stellar
colors to photospheric abundances over a wide wavelength baseline. The
clear advantage of using a photometric metallicity technique is the
efficiency of estimating metallicities for individual main-sequence
stars, which are the most plentiful and representative sample of
stellar populations.

\cite[Ivezi{\'c} et al. (2008a)]{Ive08a} constructed photometric metallicity relations in the
$u-g$ vs. $g-r$ plane using SDSS filter passbands, and studied the abundance
structures of the Galaxy with an accuracy of $\sim 0.2$ dex at $g <
17$. This approach is similar to the traditional $UBV$ method
(e.g., \cite[Carney 1979]{Car79}), which relies on the strong dependence of $U$-band
magnitudes on metal abundance. However, the $u$-band photometry in
SDSS is limited to $u \approx$ 22 (99\% detection limit). This, and
the greatly deteriorating errors in SDSS $u$-band magnitudes near the
faint limit, restricts photometric metallicity estimates to stars with
$r \lesssim 20.8$, an insufficient depth to fully explore the VOD.
Here we overcome the limitations of the $u$-band photometry in SDSS by exploring
less metallicity-sensitive, but better-determined, color indices in
the $gri$ passbands.  A full description of this method and our results can
be found in \cite[An et al. (2009b)]{An09b}.  Here we only discuss the highlights.

\section{Photometric Metallicity Estimates}

The set of stellar isochrones in \cite[An et al. (2009a)]{An09a} were
used to determine photometric metallicity estimates based on
color-color relations for main-sequence stars. A photometric
metallicity (${\rm [Fe/H]_{phot}}$) was computed for each star by
requiring distances from main-sequence fitting to be the same from two
different CMDs, with $g - r$ and $g - i$ as color indices and $r$ as a
luminosity index. We searched the entire [Fe/H] grid from $-3.0$
to $+0.4$ to find [Fe/H] with a minimum $\chi^2$ value, where we
defined

\begin{equation}
\chi^2 = \frac{(\mu_{g-r} - \bar{\mu})^2}{\sigma_{\mu_{g-r}}^2}
          + \frac{(\mu_{g-i} - \bar{\mu})^2}{\sigma_{\mu_{g-i}}^2},
\end{equation}
for each star. Here, $\mu$ and $\sigma_{\mu}$ are the distance modulus and its
error on each CMD, respectively. The $\bar{\mu}$ is a weighted average distance
modulus from the $(g - r, r)$ and $(g - i, r)$ CMDs. In principle, this method
can be used for multiple color indices. However, since only three passbands are
considered here, the problem is reduced to the traditional manner of determining
metal abundances from a color-color diagram. In some of the cases, a star
becomes bluer than the main-sequence turnoff as the metallicity increases,
either due to a large photometric error or a younger age than our assumed values
in the models. If the minimum $\chi^2$ was not found, ${\rm [Fe/H]_{phot}}$ was
estimated by extrapolating the difference in distance modulus toward higher
${\rm [Fe/H]}$ values based on theoretical predictions.

Although the accuracy of an individual photometric metallicity
estimate for a star obtained with this technique is rather large (0.5
dex or more), when the method is applied to stellar populations,
random errors are beaten down by the large numbers of stars involved. 
In the VOD, for instance (see Figure 1), each pixel in the map is
occupied by over 500 stars. The median metallicity of stars in the VOD
area at $270^o \le l \le 330^o$ and $60^o \le b \le
70^o$ is [Fe/H] $= -2.0 \pm 0.1$ from the metallicity map in
Figure 1, where the error is from a pixel-to-pixel dispersion.

\begin{figure}[t]
\begin{center}
\includegraphics[width=2.52in]{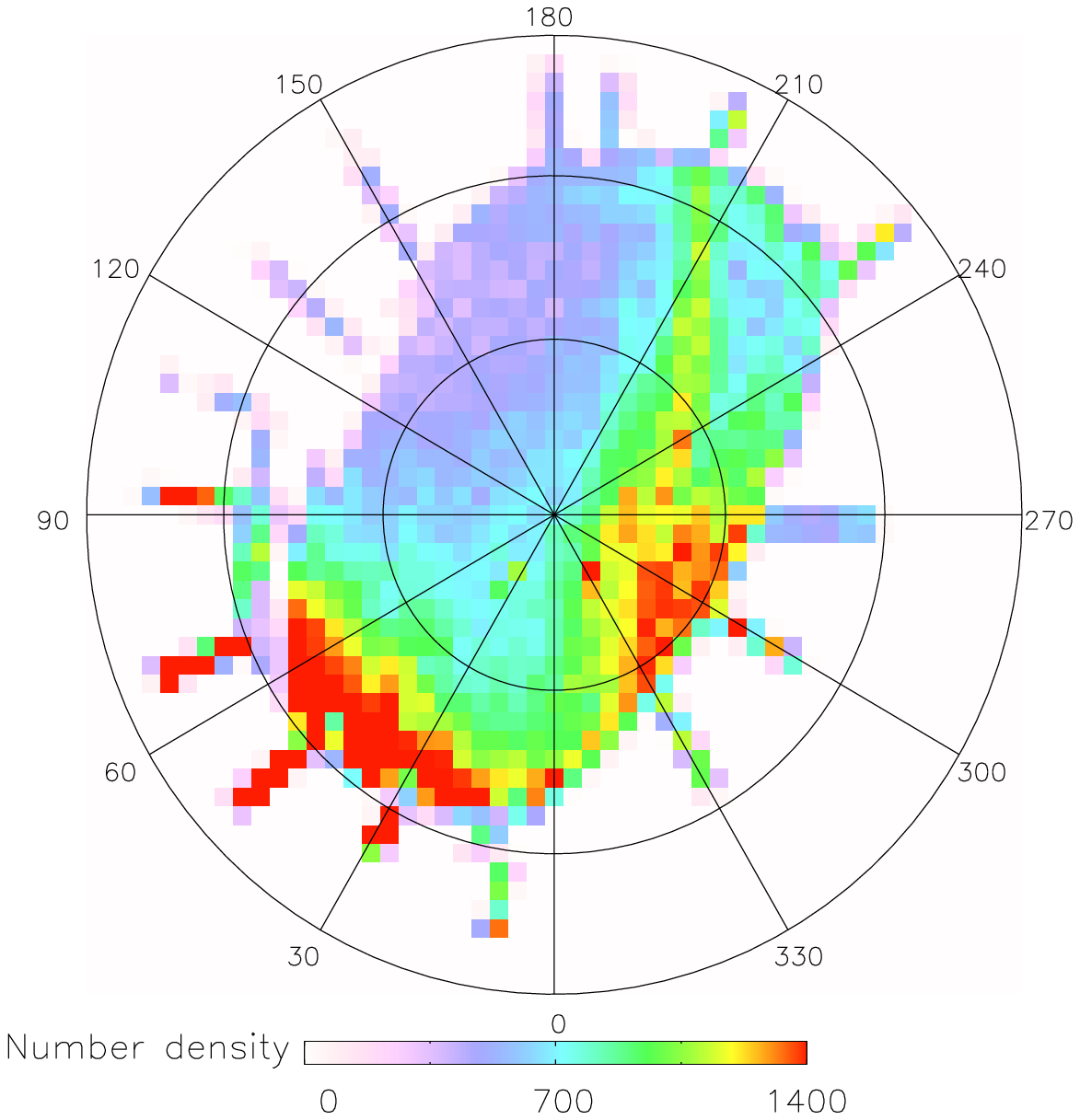}
\includegraphics[width=2.52in]{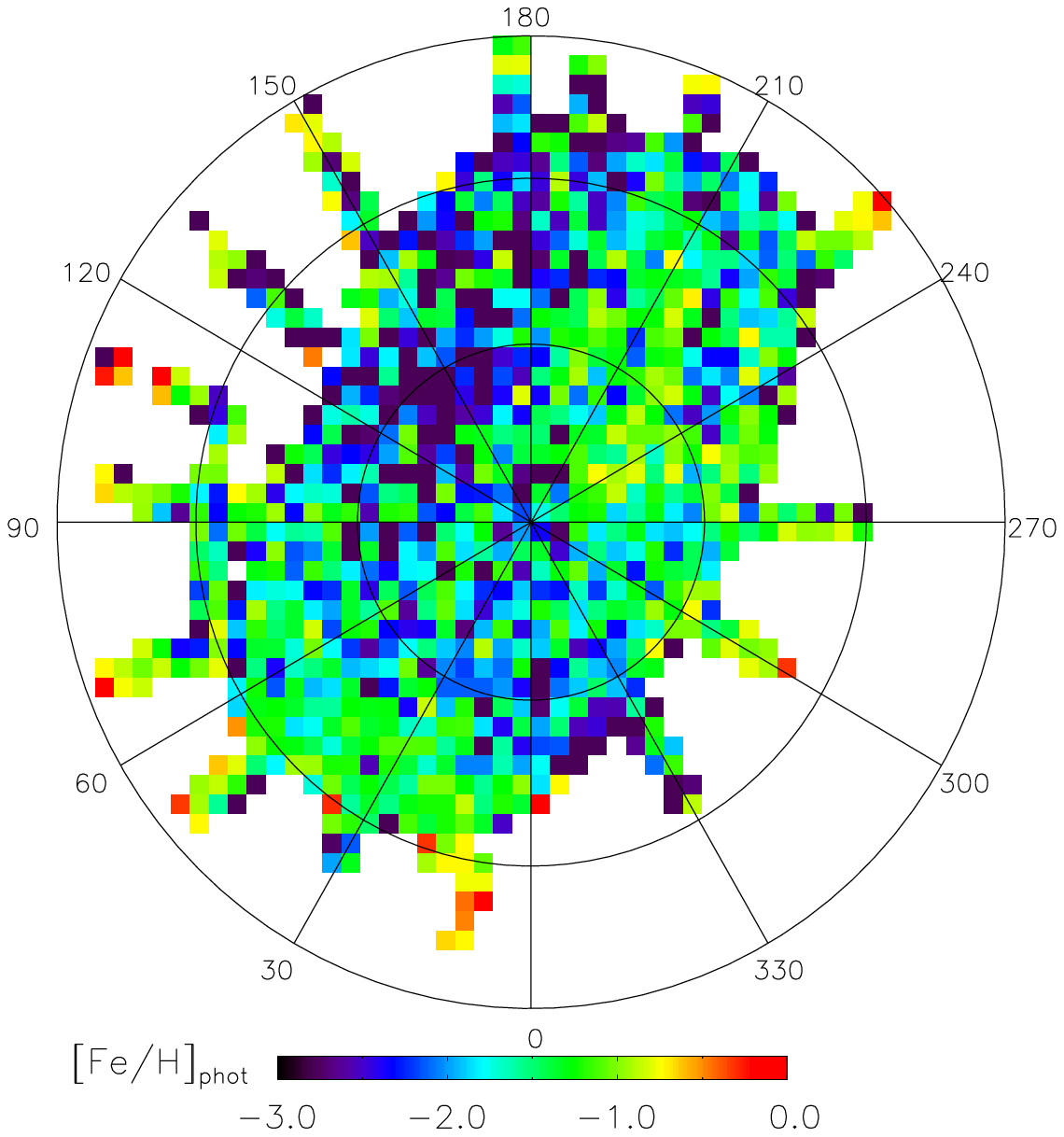} 
\caption{{\it Left}: Number density for over 740,000 stars detected in
the SDSS, located at distances from $\sim10$~kpc to $\sim20$~kpc
from the Sun in the Lambert projection of Galactic coordinates. 
Each pixel has an area of 12.96 deg$^2$, and have a median occupancy
of 544 stars per pixel. The
North Galactic Pole is at the center, and the Galactic Center is at
the bottom. Concentric circles represent $b = 0^o$, $30^o$,
and $60^o$, respectively. The Virgo Overdensity is the feature
seen at $(l,b) \sim (300^o,70^o)$, while the long strip
that crosses the sky from $(l,b) \sim (210^o,30^o)$ to the
VOD is the leading tidal tail of Sgr. The feature at $(l,b) \sim
(40^o,40^o)$ is the Hercules-Aquila Cloud, another large-area overdensity of
halo stars discovered in SDSS (\cite[Belokurov et al. 2007]{Bel07}). {\it Right}:
Median metallicity of the same stars as in the {\it Left} panel.}
\label{fig1}
\end{center}
\end{figure}

Field halo stars in the mirrored position of the map shown in Figure 1
at $30^o \leq l \leq 90^o$ and $60^o \leq b \leq
70^o$ exhibit ${\rm \langle [Fe/H] \rangle} = -1.9 \pm 0.1$.
Although half of the stars in the direction toward Virgo are likely
associated with a progenitor dwarf galaxy or a tidal stream, the mean
[Fe/H] value essentially remains unchanged; it is the same as that for
the field halo stars within the precision of the technique. It is
perhaps of interest that the field star metallicities are as low as
they appear to be, as \cite[Carollo et al. (2007)]{Car07} have argued
that the peak of the metallicity of the outer-halo population is
[Fe/H]$ = -2.2$, and that this component is expected to dominate over
the more metal-rich inner-halo population (with a peak metallicity at
[Fe/H]$ = -1.6$) at Galactocentric distances greater than 15-20 kpc. 

\section{A Metallicity Map of the Halo to 20 kpc}

The same techniques described above can be used to construct a
metallicity map, based on the $gri$ passbands, for main-sequence stars
in the SDSS footprint extending well beyond the limit of 9 kpc reached
by the metallicity map presented by \cite[Ivezi{\'c} et al.
(2008a)]{Ive08a}, based on the $ugr$ passbands. A preliminary result 
of this exercise is shown in
Figure 2. We are still in the process of more fully investigating the
effect of photometric errors at low metallicity, however, the current
figure certainly suggests that the typical
stellar metallicity inferred for stars at larger distances is
substantially lower than that for more nearby stars. 

\cite[Carollo et al. (2009)]{Car09} have
recently used a new, much larger sample of calibration stars from SDSS
DR-7 to develop a more quantitative picture of the
inner/outer halo dichotomy. Their derived velocity ellipsoids for
these two components -- Inner Halo: ($\sigma_{V_{R}}$,
$\sigma_{V_{\phi}}$, $\sigma_{V_{Z}}$) = (150 $\pm$ 2, 100 $\pm$ 2, 85
$\pm$ 1) km~s$^{-1}$, Outer Halo: ($\sigma_{V_{R}}$,
$\sigma_{V_{\phi}}$, $\sigma_{V_{Z}}$) = (159 $\pm$ 4, 165 $\pm$ 9,
116 $\pm$ 3) km~s$^{-1}$, clearly demonstrate that outer-halo orbits
are significantly less radially elongated than inner-halo orbits, and
possess higher energies, consistent with the segregation in spatial
dimensions observed in the metallicity map.

\begin{figure}[t]
\begin{center}
 \includegraphics[width=2.6in]{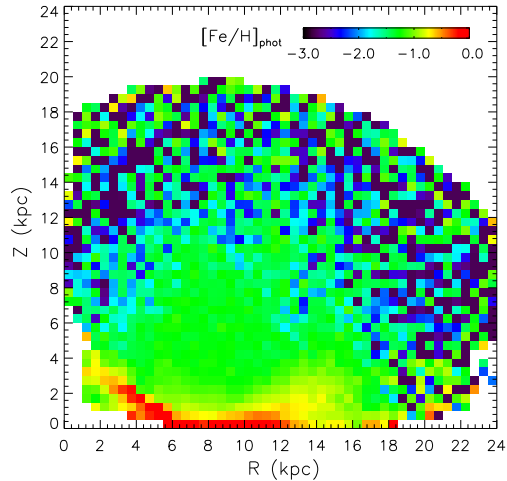} 
 \caption{Metallicity map, obtained from photometric metallicity
estimates for 2.9 million main-sequence stars with $0.3 < g-r < 0.4$
in the northern SDSS DR-7 footprint.  The median
occupancy for pixels shown in the map is 638 stars per pixel. 
Note that the median metallicity for stars located beyond 10 kpc
appears much lower than those within 10 kpc.} \label{fig2}
\end{center}
\end{figure}

Future imaging surveys, such as Pan-STARRS (\cite[Kaiser et al.
2002]{kai02}) and LSST (\cite[Ivezi{\'c} et al. 2008b]{Ive08b}) will use
similar photometric bandpasses as those in SDSS, providing even deeper
(and {\it far} more accurate) photometric data than SDSS over a larger
fraction of the sky. Our photometric metallicity method will be useful
to exploit these databases for understanding the chemical evolution of
progenitor dwarf galaxies that are identified, as well as for the bulk
populations of field stars.
 
{}

\end{document}